\documentclass[12pt]{article}
\usepackage{amsmath,amsfonts}
\makeatletter \@addtoreset{equation}{section}

\usepackage{cite}
\usepackage{bm}
\usepackage{dcolumn}
\usepackage{graphicx}
\usepackage{subfig}
\usepackage{float}
\makeatother
\def\one{{\hbox{ 1\kern-.8mm l}}}
\newcommand{\Dslash}{\not{\hbox{\kern-4pt $D$}}}
\newcommand{\pdslash}{\not{\hbox{\kern-2pt $\partial$}}}
\newcommand{\be}{\begin{equation}}
\newcommand{\bea}{\begin{eqnarray}}
\newcommand{\eea}{\end{eqnarray}}
\newcommand{\ba}{\begin{array}}
\newcommand{\ea}{\end{array}}
\newcommand{\ee}{\end{equation}}

\newcommand{\bss}{\begin{split}}
\newcommand{\ess}{\end{split}}

\begin{document}

\begin{titlepage}
\vspace*{1mm}%
\hfill%
\vbox{
    \halign{#\hfil        \cr
           IPM/P-2010/026 \cr
                     } 
      }  
\vspace*{15mm}%
\begin{center}

{{\Large {\bf Holographic Phase Transition \\ \vspace*{5mm} to\\
\vspace*{5mm}%
 Topological Dyons
}}}

\vspace*{15mm} \vspace*{1mm}{Davood Allahbakhshi$^{a,b}$ and
Farhad Ardalan$^{a,b}$}

 \vspace*{1cm}

{\it $^a$ School of physics, Institute for Research in Fundamental Sciences (IPM)\\
P.O. Box 19395-5531, Tehran, Iran \\}

\vspace*{.4cm}

{\it $^b$ Department of Physics, Sharif University of Technology \\
P.O. Box 11365-9161, Tehran, Iran}

\vspace*{.4cm}

E-mails: \it{allahbakhshi@ipm.ir , ardalan@ipm.ir}

\vspace*{2cm}
\end{center}

\begin{abstract}
The dynamical stability of a Julia-Zee solution in the AdS
background in a four dimensional Einstein-Yang-Mills-Higgs theory
is studied. We find that the model with a vanishing scalar field
develops a non-zero value for the field at a certain critical
temperature which corresponds to a topological dyon in the bulk
and a topological phase transition at the boundary.

\end{abstract}

\end{titlepage}


\section{Introduction}
The gravity-gauge theory duality, AdS/CFT
\cite{Maldacena:1997re}\cite{Witten:1998qj}\cite{Gubser:1998bc},
has provided the hope of doing quantitative analysis in low energy
QCD. It is suspected that the phenomenon of confinement  in QCD
may be related to the condensation of topological configurations,
specifically magnetic monopoles in QCD
\cite{Nambu:1974zg}\cite{Mandelstam:1974pi}\cite{Polyakov:1976fu}
. Extended  topological configurations in field theories have
other significant consequences.

In this work we study the consequence of one such topological
 charged magnetic configuration, a dyon, in the bulk of an AdS/ CFT
set up. The configuration being the dyon solution of an SU(2)
gauge theory coupled to a triplet of scalar fields, of
'tHooft-Polyakov-Julia-Zee type, in the asymptotically AdS
Reissner-Nordstrom blackhole background.  We will argue that this
will emerge in the low temperature regime from a configuration
with no scalar field. The boundary theory will make a
corresponding phase transition.

Magnetic monopoles have a long history. The gauge theory in 4
dimensions, the electrodynamics, does not allow magnetic charges
at the right hand side of the Maxwell  equation for the dual field
intensity vanishes identically, unless a singularity, the Dirac
string, is inserted. A similar situation in non-abelian gauge
theories persists.

Nearly thirty five years ago Polyakov and 'tHooft observed that
introduction of an adjoint scalar  permits the appearance of a
magnetic monopole solution in the SU(2) gauge theory, upon
spontaneous breaking of the gauge symmetry to a U(1) subgroup
determined by the direction of the scalar field in isospin space.
The magnetic charge being a topological quantity, is classically
quantized \cite{'tHooft:1974qc}\cite{Polyakov:1974ek}.

This "Hedge-hog" solution led to an avalanche of work on
generalization to other gauge groups with serious problems for
cosmology, whose resolution was achieved in the theory of
inflation.

Shortly after the discovery of the solutions with magnetic charge,
dyons with electric charge also were found by Julia and Zee
\cite{Julia:1975ff}. A particular limit in the solutions
discovered by Prasad and Sommerfield \cite{Prasad:1975kr},
saturating a bound observed by Bogomol'nyi
\cite{Bogomolny:1975de},the BPS solution has permeated the
literature of supergravity and string theory.

Magnetic monopole solutions in the presence of gravity have also
attracted a great deal of attention. Many solutions  in
asymptotically flat, deSitter, or  Anti-deSitter spaces with and
without blackhole singularity have been discovered
\cite{Volkov:1998cc}. Remarkably only the ones with the
Anti-deSitter asymptotics are stable
\cite{Bjoraker:2000qd}\cite{Mosaffa:2006gp}. This is a striking
result in view of  the AdS/CFT duality.

In particular for SU(2) gauge theory, asymptotically AdS blackhole
solutions with and without scalar field have been discovered
\cite{Winstanley:1998sn}, and more recently by Lugo, Moreno, and
Schaposnik,\cite{Lugo:2010qq}.

We will use a particularly simple solution found quite some time
ago by Kasuya and Kamata \cite{Kasuya:1981tq}, in the present
work. We will find that their solution in the absence of a scalar
field is unstable while the one with non-vanishing scalar triplet
is.

The paper is organized as follows. In the next section we briefly
review the Julia-Zee dyon in flat space. In section 3 we introduce
it's extension to gravitational versions. In section 4 we review
the general procedure for studying dynamical stability. In section
5 we apply this procedure to our case and derive the main analytic
equations. In section 6 we present the results of our numerical
calculations. Section 7 is devoted to conclusion.

When this work was completed, a paper by Lugo, Moreno, and
Schaposnik  appeared \cite{Lugo:2010ch}, with similar
considerations, related to their previous solution and in the
context of the noncompact boundary of $R^2$ in place of $S^2$.

\section{Julia Zee dyon in flat space}

In 1974 'tHooft\cite{'tHooft:1974qc} and
Polyakov\cite{Polyakov:1974ek} independently introduced a new type
of magnetic monopole in a flat Minkowski space. Their monopole is
free of any singularities. The action of the theory is
Yang-Mills-Higgs with a special nontrivial ansatz for the gauge
and scalar fields. The action is: \be S \ = \int d^4x\
[-\frac{1}{4}F^{a}_{\mu \nu}F^{a\mu\nu}-\frac{1}{2}D_\mu\phi ^a
D^\mu \phi ^a \ -\ V(\phi^a\phi^a)], \ee

The scalar potential is assumed to have a minimum at
$\phi^a\phi^a=(\vec{\phi})^2 = const\neq 0$. The covariant
derivative and the field strength are defined as usual,

\be D_\mu\phi^a=\partial_\mu \phi^a + e \epsilon^{abc}A^b_\mu
\phi^c\;;\;F^a_{\mu\nu} = \partial_\mu A^a_\nu - \partial_\nu
A^a_\mu+ e
\epsilon^{abc}A^b_\mu A^c_\nu. \ee\\
The 'tHooft-Polyakov ansatz for a solution for this system is,
\be\vec{\phi}=\frac{H(r)}{er}\hat{r} \ ;\
\vec{A}_i=\frac{1-k(r)}{er}\vec{a}_i\ (i=1,2,3)\ ; \
a^n_i=\epsilon^{nik}\hat{x}^k \ ; \vec{A}_t=0\ee

Regularity and finiteness of energy requires the boundary conditions,\\\\
at $ r \rightarrow 0\ : \ H\rightarrow0 \ ;K \rightarrow 1 $\\at $
r\rightarrow\infty \ : \ H \rightarrow const \times r \ ; \
K\rightarrow 0$ \\\\The U(1) field strength, invariant under the
su(2) algebra, is defined as,\be
F_{\mu\nu}=\vec{F}_{\mu\nu}.\hat{\phi}+\hat{\phi}.[D_\mu\hat{\phi}\times
D_\nu\hat{\phi}]\;;\;\hat{\phi}=\frac{\vec{\phi}}{\sqrt{\vec{\phi}.\vec{\phi}}}\;;\ee
When $r\rightarrow\infty$, the magnetic field scales as $1/r^2$;
so the configuration has a magnetic charge:\be \lim_{r \to \infty}
B_i\to\frac{1}{e \; r^2}\Rightarrow \oint{B_i\;ds^i}=\frac{1}{e}=g
\ee Later Arafune, Freund, and Goebel \cite{Arafune:1975ab} showed
that this magnetic charge is a topological object and its value is
discrete,\be g=\oint{B_i\;ds^i}=\frac{n}{e}. \ee

The hamiltonian of the system can be written as, \be H=\int d^3x\
[\frac{1}{4}\big(F^{a}_{ij}-\epsilon_{ijk} D_k\phi ^a\big)^2 +
V(\phi^a\phi^a)]+4\pi g \langle\phi\rangle , \ee where
$\langle\phi\rangle$ is the boundary value of $\phi$ which is a
constant by assumption. It is then easily found
\cite{Bogomolny:1975de} that there is a bound on the energy, BPS
bound,\be E\geq 4\pi g \langle \phi\rangle.\ee The bound is
saturated when \be F_{ij}^a=\epsilon_{ijk}D_k\phi^a.\ee

This magnetic monopole solution was later extended to a dyon by
Julia and Zee\cite{Julia:1975ff}. Their solution was in the form:

\bea\begin{split}\vec{\phi}&=\frac{H(r)}{er}\hat{r}, \cr
\vec{A}_\alpha &= \frac{1-K(r)}{e}\vec{a}_\alpha, \cr \vec{A}_t &=
\frac{J(r)}{er}\hat{r}.\end{split}\eea

The boundary conditions for $H$ and $K$ are as before, and the
boundary conditions for $J(r)$ are:\\\\at $ r \rightarrow 0\ : \
J\rightarrow0 \ $\\at $ r\rightarrow\infty \ : \ J \rightarrow
const \times r$\\\\With this ansatz we see that the configuration
has an electric charge too.

In 1975 Prasad and Sommerfield\cite{Prasad:1975kr} found an exact
analytic solution for this dyon in the limit $V \to 0$. their
solution has the form:\bea \begin{split}K&=\frac{Cr}{sinh(Cr)},\cr
J&=sinh(\gamma)\big[Cr\;coth(Cr)-1\big],\cr
H&=cosh(\gamma)\big[Cr\;coth(Cr)-1\big],\end{split}\eea where $C$
and $\gamma$ are arbitrary constants.

\section{Julia-Zee dyon coupled to gravity}

The dyon of the previous section lives in flat space but what
happens if we include the gravity? During 80's it was shown that
there exist gravitational Julia-Zee dyons; and certain exact
solutions were obtained \cite{Comtet:1979kq}. One of the simplest
solutions in this class is the Kasuya and Kamata
solution\cite{Kasuya:1981tq}.

When gravity is included the action is:\be S \ = \int \sqrt{-g}\;
d^4x\ \big[\frac{1}{16 \pi
G}(R+\frac{6}{L^2})-\frac{1}{4}F^{a}_{\mu
\nu}F^{a\mu\nu}-\frac{1}{2}D_\mu\phi ^a D^\mu \phi ^a \ -\lambda\
V(\phi^a\phi^a)\big]\ee And as before the scalar potential is
assumed to have a minimum at some nonzero constant
$(\vec{\phi})^2$. The equations of motion are: \be \begin{split}
&R_{\mu\nu}-\frac{1}{2}(R+\frac{6}{L^2})\;g_{\mu\nu}=8\pi G
\;T_{\mu\nu},\cr &T_{\mu\nu}=\big[\frac{\Lambda}{16\pi G}
-\frac{1}{4}F_{\rho\lambda}^a
F^{a\rho\lambda}-\frac{1}{2}D_\lambda\phi^a
D^\lambda\phi^a-\lambda V(\phi^2) \big]
g_{\mu\nu}+F^a_{\mu\lambda}F^{a\lambda}_\nu+D_\mu\phi^a
D_\nu\phi^a,\cr\cr &\partial_\mu\big( \sqrt{-g}D^\mu\phi^a
\big)+\sqrt{-g}\big(e \epsilon^{abc}A^b_\mu
D^\mu\phi^c-\lambda\;\frac{\delta V}{\delta \phi^a} \big)=0,
\cr\cr &\partial_\mu\big( \sqrt{-g}F^{a\mu\nu}
\big)-e\sqrt{-g}\;\epsilon^{abc}\big(F^{b\mu\nu}A^c_\mu+(\partial^\mu\phi^b)\;\phi^c\big)+e^2\sqrt{-g}\big((A^{b\nu}
\phi^b)\phi^a-(\phi^b\phi^b) A^{a\nu}\big)=0.
\end{split} \ee

The same ansatz for scalar and vector fields is assumed as in the
flat space case. The most general form for a spherically symmetric
metric in 4 dimensions is: \be g_{\mu\nu}\equiv
diag(-e^{X(r)},e^{Y(r)},r^2,r^2sin(\theta)^2)\ee If the scalar
potential is taken to be in the form of a mexican hat,
$V\;=\;\frac{1}{4}(\phi^a\phi^a)^2-\frac{v^2}{2}(\phi^a\phi^a)$,
then the equtions of motion for $H,K$ and $J$ become:\bea
\begin{split}&J^{\prime\prime}-\frac{r}{2}(X+Y)^{\prime}\big(\frac{J}{r}\big)^\prime-e^Y\frac{2JK^2}{r^2}=0,
\cr &K^{\prime\prime}+\frac{1}{2}(X-Y)^\prime
K^\prime-e^Y\frac{K}{r^2}\big(K^2+H^2-e^{-X}J^2-1\big)=0, \cr
&H^{\prime\prime}+\frac{r}{2}(X-Y)^\prime\big(\frac{H}{r}\big)^\prime-e^Y\frac{H}{r^2}\big(2K^2+\frac{\lambda}{e^2}(H^2-c^2r^2)\big)=0,\end{split}
\eea
where $c^2=e^2v^2$ and $v$ is the minimum of the scalar potential.\\
The equations of motion for the metric yield,\bea
\begin{split}&\frac{e^{-Y}}{r^2}(rY^\prime-1)+\frac{1}{r^2}=\frac{8\pi
G}{e^2}T^t_t, \cr
-&\frac{e^{-Y}}{r^2}(rX^\prime+1)+\frac{1}{r^2}=\frac{8\pi
G}{e^2}T^r_r, \cr
-&\frac{e^{-Y}}{2}\big[X^{\prime\prime}+\frac{1}{2}(X^\prime)^2-\frac{1}{2}X^\prime
Y^\prime+\frac{1}{r}(X-Y)^\prime\big]=\frac{8\pi
G}{e^2}T^\theta_\theta=\frac{8\pi
G}{e^2}T^\varphi_\varphi,\end{split}\eea where
$T^t_t,T^r_r,T^\theta_\theta$ and $T^\varphi_\varphi$ are the
components of the energy momentum tensor, \bea
\begin{split}
&T^t_t\;=\;\bigg[\frac{e^{-Y}}{r^2}(K^\prime)^2+\frac{(K^2-1)^2}{2r^4}+\frac{e^{-(X+Y)}}{2}[(\frac{J}{r})^\prime]^2+\cr
&e^{-X}\frac{J^2K^2}{r^4}+\frac{e^{-Y}}{2}[(\frac{H}{r})^\prime]^2+\frac{H^2K^2}{r^4}+e^2\lambda
V(\phi)\bigg],\cr\cr
&T^r_r\;=\;\bigg[-\frac{e^{-Y}}{r^2}(K^\prime)^2+\frac{(K^2-1)^2}{2r^4}+\frac{e^{-(X+Y)}}{2}[(\frac{J}{r})^\prime]^2-\cr
&e^{-X}\frac{J^2K^2}{r^4}-\frac{e^{-Y}}{2}[(\frac{H}{r})^\prime]^2+\frac{H^2K^2}{r^4}+e^2\lambda
V(\phi)\bigg],\cr\cr
&T^\theta_\theta\;=\;T^\varphi_\varphi\;=\;\bigg[-\frac{(K^2-1)^2}{2r^4}-\frac{e^{-(X+Y)}}{2}[(\frac{J}{r})^\prime]^2+\frac{e^{-Y}}{2}[(\frac{H}{r})^\prime]^2+e^2\lambda
V(\phi)\bigg].
\end{split}\eea
Kasuya and Kamata found an exact solution to these equations with
the above ansatz \cite{Kasuya:1981tq}:\bea\begin{split} H=c\;r \;
; \; K=0\; ; \; J=\mu\;r-\rho\; ; \;
e^{X}=e^{-Y}=1-\frac{2M}{r}+\frac{q^2}{r^2}+\frac{r^2}{L^2},
\end{split}\eea
where,\be q^2=\frac{4\pi G_N(1+\rho^2)}{e^2}.\ee

This expression is still a solution if we set $H=0$. The
background metric in both cases are \textit{AdS-RN}; but, there is
an important point : The AdS radius differs for the two, because
the contribution of the scalar potential to the cosmological
constant is different for the two cases. Note also that because
the gauge field should have a finite norm, $A^a_t$ should be zero
at the horizon of the blackhole, implying a relation between $\mu$
and $\rho$ \be \mu \;=\;\frac{\rho}{r_H}\;,\ee where $r_H$ is the
outer horizon of the blackhole. The temperature of the blackhole
after eliminating the charge of the blackhole from definition of
the horizon ($e^X=0$) is:\be T=\frac{1}{2\pi
r_H}\big(1-\frac{M}{r_H}+\frac{2r_H^2}{L^2}\big). \ee

Concerning the Bogomol'nyi equation, when the gravity is included,
the simplest guess is: \be
D_i\phi^a\;=\frac{1}{2}\sqrt{-g}\;\epsilon_{ijk}F^{ajk}. \ee But
this does not work; in fact this relation is not compatible with
the equations of motion derived from the action (3.1). It has been
shown that the action of the theory can be changed in such a way
that its equations of motion are compatible with the relation
(3.11)\cite{Zhang:2009tt}. On the other hand in certain
circumstances a Bogomol'nyi like equation can be written which is
compatible with the action (3.1). For example one may consider a
generalization of the Bogomol'nyi equation (2.9) of the form,\be
D_i\phi^a\;+\;\partial_i
(u)\;\phi^a=\frac{1}{2}\sqrt{-g}\;\epsilon_{ijk}F^{ajk}, \ee where
u is an additional function. Consistency of this equation with the
equations of motion leads to:\be
u=log(\sqrt{|g_{00}|}\;)\;;\;\Delta u=0, \ee where $\Delta$ is the
usual covariant Laplacian; this in fact is a constraint on the
metric. For more details see\cite{Comtet:1983wt}.

\section{General aspects of dynamical stability}
In this section we review the general aspects of dynamical
stability analysis and holographic phase transition.

In the presence of a background solution for a system, checking
dynamical stability is to study the time evolution of the
solution. Consider a system with a set of fields $X_i(x)$ and the
equations of motion $E_i[X_j(x)]=0$, and a set of exact solutions
$X^{(0)}_i(x)$. Varying the solution with an infinitesimal time
dependent variation, the linearized equation will be, \be
L_i[X^{(0)}_j,\delta \tilde{X}_j(x),\omega]=0,\ee with $\delta
X_i(x,t)=\delta \tilde{X}_i(x)\;e^{-i\omega t}$.

If these equations admit a solution with an $\omega$ which has
positive imaginary part, then the variations blow up with time and
the background solution will be unstable.

This simple argument is the core of the concept of dynamical
instability, but there are a number of technical issues concerning
the boundary conditions on the fields which are important:

The variations should be regular at any point of the space, as we
assumed these variations to be infinitesimal.

Also, at the horizon because of the general properties of the
non-extremal blackholes, the radial dependence of the variations
 near the horizon is generally of the form
\cite{Gubser:2008wv},\be \delta \tilde{X}_i(r)=(r-r_H)^{\pm
i\frac{\omega}{4\pi
T}}\big(a_0+a_1(r-r_H)+a_2(r-r_H)^2+...\big),\ee where T is the
temperature of the blackhole, and the $\pm$ signs indicate the
ingoing and outcoming modes. Then as a classical blackhole devours
everything, the solution near the horizon should be ingoing only.

Moreover, at the boundary at infinity, because the equations are
of order two, the solution asymptotically has the form,\be
\delta\tilde{X}\sim\;c_1\;r^{\Delta_+}\;+\;c_2\;r^{\Delta_-}\;.
\ee There are then two possibilities:

1) One of the modes diverges (non-normalizable); the diverging
mode must be excluded as the variation should remain small; while
the other one vanishes at the boundary (normalizable) .

2) Both modes vanish. Then the one that is coupled to the
appropriate boundary operator should be chosen. In the context of
AdS/CFT such normalizable modes are interpreted as the v.e.v of
the corresponding operator at the field theory (CFT) side.

So our boundary conditions are: \bea\begin{split} &\text{near
horizon}&&\rightarrow \text{ingoing mode},\cr &\text{at the
boundary} &&\rightarrow \;
c_1\;r^{\Delta_+}\;+\;c_2\;r^{\Delta_-}.
\end{split}\eea
The coefficients $c_1$ and $c_2$ are functions of parameters in
the theory, e.g. temperature, radius of the blackhole, and
chemical potential. Then choosing one of the boundary modes, e.g.
$c_1=0$, leads to the allowed frequencies: \be
c_1(\;T,\omega,r_H,\mu,...)=0\;\rightarrow\;\omega(\;T,r_H,\mu,...).\ee

Then if the imaginary part of the $\omega$ changes sign in the
vicinity of a hypersurface (\emph{wall of marginal stability}) in
the ( $T,r_H,\mu,...$) space, then there will be a
\textit{stability/instability} transition for the configuration,
and the system goes through a phase transition.

The modes on the wall of marginal stability with $\omega$ equal to
zero are called \emph{marginally stable
modes}\cite{Gubser:2008px}.

\section{Stability analysis of gravitational Julia-Zee like solution}
We consider the quartic potential:\be \lambda
 V(\phi^a\phi^a)=\frac{\lambda}{4}(\phi^a\phi^a)^2-
 \frac{1}{L^2}(\phi^a\phi^a).\ee
As mentioned in section 3, there are two exact solutions to the
equations of motion for Julia-Zee ansatz in the form of (2.10):
$H=0$ with AdS radius L; and $H=\;r \; \sqrt{\frac{2e^2}{\lambda
L^2}}$ with AdS radius $\tilde{L}$ :\be
\frac{1}{\tilde{L}^2}=\frac{1}{L^2}+\frac{8\pi G_N}{3\lambda L^4}.
\ee Perturbing the matter fields,\bea
\begin{split} &\frac{H}{er} \rightarrow \frac{H}{er}+\epsilon \;
e^{-i\omega\; t}\;\frac{f(r)}{e},\cr &\frac{J}{er}\rightarrow
\frac{J}{er}+\epsilon \; e^{-i\omega\; t}\;\frac{P(r)}{e},\cr
&\frac{K}{e}\rightarrow \frac{K}{e}+\epsilon \; e^{-i\omega\;
t}\;\frac{Q(r)}{e},
\end{split} \eea
where $\epsilon$ is an infinitesimal parameter, and putting the
new fields in the equations of motion for the solution $H=0$, to
first order in $\epsilon$, we get:\bea
\begin{split} &\phi \; &&equation:
\;\big(r^2e^X\;f'\big)'+\big(\frac{2}{L^2}+e^{-X}\omega^2\big)\;r^2\;f=0,\cr
&A_t \; &&equation:\;2P'+r\;P''=0,\cr &A_r\;&&equation:\;P'=0,\cr
&A_\theta\;&&equation:\;\left\{
                          \ba{ll}
                            & \hbox{$Q''+X'Q'+e^{-2X}\big[e^X/r^2+(\mu-\rho/r)^2+\omega^2\big]Q=0 $}, \\
                            & \hbox{$Q=0$},
                          \ea
                        \right.
\cr &A_\varphi\; &&equation:\;\left\{
                          \ba{ll}
                            & \hbox{$Q''+X'Q'+e^{-2X}\big[e^X/r^2+(\mu-\rho/r)^2+\omega^2\big]Q=0 $}, \\
                            & \hbox{$Q=0$}.
                          \ea
                        \right.
\end{split}\eea

The solutions to the gauge field perturbations are simply
$P=const\;,\;Q=0$; but, the equation for the variation of $\phi$
is nontrivial. We can consider different perturbations to
$A_\varphi$ and $A_\theta$, but the $A_t$ equation implies that
they should be equal.

The first step is to determine the boundary behavior of the
function $f$. The near horizon equation for $f$ is:\be
f''+\frac{f'}{u}+\big(\frac{\tilde{\eta}}{u}+\frac{\tilde{\omega}^2}{u^2}\big)f=0,\ee
where $u$ is $(r-r_H)$ , and,\be \tilde{\eta}=\frac{1}{2\pi T
L^2}\; ; \; \tilde{\omega}=\frac{\omega}{4\pi T},\ee with the
solution,\be J_{\pm
2i\sqrt{\tilde{\omega}^2}}(2\sqrt{\tilde{\eta}\;u}), \ee which has
the small u
 expansion,\be u^{\pm i \sqrt{\tilde{\omega}^2}} \;\big(\;\frac{\tilde{\eta
}^{\pm i \sqrt{\tilde{\omega}^2}}}{\Gamma[1\pm 2 i
\sqrt{\tilde{\omega}^2}]}\;-\;\frac{\tilde{\eta }^{1\pm i
\sqrt{\tilde{\omega}^2}}}{\Gamma[2\pm 2 i
\sqrt{\tilde{\omega}^2}]}\;u\;+\;\frac{\tilde{\eta }^{2\pm i
\sqrt{\tilde{\omega}^2}}}{2 \Gamma[3\pm 2
i\sqrt{\tilde{\omega}^2}]}\;u^2\;+\;O[u]^3\;\big). \ee This
corresponds to the form (4.2).

The boundary equation for $f$ at $r\rightarrow \infty$ is :\be
f''+\frac{4}{r}f'+\big(\frac{2}{r^2}+\frac{L^4\omega^2}{r^4}\big)f=0.
\ee Changing the variable to $z=1/r$,\be
f^{\prime\prime}-\frac{2}{z}f^{\prime}+\big(\frac{2}{z^2}\;+\;L^4\omega^2\big)f=0,
\ee which has the solution,\be f\;=\;z\;e^{\pm i L^2\omega
\;z}=\frac{e^{\pm i \frac{L^2\omega}{r}}}{r}. \ee For large $r$
this has the form, \be f\sim\frac{f_1}{r}+\frac{f_2}{r^2}+... \ee

As described in the last section a choice has to be made; at the
horizon the ingoing mode is to be chosen and at the boundary at
infinity either $f_1$ or $f_2$ should be set equal to zero. Each
 leads to a different value for the critical temperature.

We can repeat exactly the same procedure for the second solution
with $H=\;r \; \sqrt{\frac{2e^2}{\lambda L^2}}$. At the linear
order the equations are as before except for the $\phi$ equation,
\be
\big(r^2e^X\;f'\big)'+\big(-\frac{4}{L^2}+e^{-X}\omega^2\big)\;r^2\;f=0.\ee
The forms of this equation near the horizon and the boundary at
infinity are:\bea
\begin{split}
f''+\frac{f'}{u}+\big(-\frac{\tilde{\eta}}{u}+\frac{\tilde{\omega}^2}{u^2}\big)f=0,
\cr f''+\frac{4}{r}f'+\big(-\frac{4\alpha^2}{r^2}+\frac{(\alpha
L)^4\omega^2}{r^4}\big)f=0,
\end{split}\eea where
\be\tilde{\eta}=\frac{1}{\pi T \tilde{L}^2}\; , \;
\tilde{\omega}=\frac{\omega}{4\pi
T}\;,\;\alpha=\frac{\tilde{L}}{L}=\frac{1}{\sqrt{1+\frac{\gamma}{L^2}}}\;,\;0\leq\alpha\leq1,\ee
 $\gamma$ is $(8\pi G_N)/(3\lambda)$, $u=r-r_h$, and " $ ' $ " in the first equation denotes derivation with
respect to $u$ and in the second equation with respect to $r$.

The solutions to the near horizon equation are:\be I_{\pm2 i
\sqrt{\tilde{\omega}^2} }(2 \sqrt{u\tilde{\eta}}),\ee which are
ingoing and outcoming modes. The solutions to the equation at the
boundary at infinity are:\be \frac{1}{r^{3/2}}J_{\pm\frac{1}{2}
\sqrt{9+16 \alpha ^2}}(\frac{L \alpha ^2 \omega }{r}),\ee with the
asymptotic expansion of the form, \bea\begin{split} f\;=
\;f_1\;r^{\Delta_+}\;+\;f_2\;r^{\Delta_-},\cr
\Delta_\pm=\frac{-3\pm\sqrt{9+16\alpha^2}}{2}.\end{split}\eea The
first term is divergent and is not acceptable, thus we set
$f_1=0$.

\section{Numerics}
We follow the procedure outlined above for finding the \emph{phase
transition} temperature by setting $\omega$ equal to zero, i.e.,
finding the marginally stable mode.
\subsection{The case $\phi=0$ :}

\subsubsection{marginally stable modes:}
The $\phi$ equation when $\omega$ is set to zero is:\be
\big(r^2e^X\;f'\big)'+\frac{2}{L^2}\;r^2\;f=0. \ee The forms of
this equation at the horizon and at the boundary at infinity
are:\bea
\begin{split} f''+\frac{f'}{u}+\frac{\tilde{\eta}}{u}f=0,\cr
f''+\frac{4}{r}f'+\frac{2}{r^2}f=0,
\end{split}\eea where $u=r-r_H$; and $" \; ' \; "$
denotes derivation with respect to $u$ in the first equation and
to $r$ in the second one.

The solutions to the first equation are:\be
J_0(2\sqrt{\tilde{\eta}\;u})\; ;
\;Y_0(2\sqrt{\tilde{\eta}\;u}).\ee Now
$Y_0(2\sqrt{\tilde{\eta}\;u})$ is divergent at the horizon so it
is ruled out; thus we choose $J_0(2\sqrt{\tilde{\eta}\;u})$, and
its near horizon expansion is:\be 1-\tilde{\eta}\;u\;+\; ... .\ee
We can fix $f$ at the horizon to be equal to 1.

The solution to the second equation is,\be f=
\frac{f_1}{r}+\frac{f_2}{r^2}. \ee Eliminating the charge of the
blackhole in favor of $M,L$ and $r_H$, then $f_1$ and $f_2$ will
be functions of ($M,L, r_H$). Imposing the desired boundary
condition, $f_1=0$ or $f_2=0$, we will find a relation between
($M,L, r_H$). Simultaneously we have the positivity condition of
the temperature , $T(M,L, r_H)\geq 0$. For convenience we fix
$r_H$ to a certain value, e.g. $r_H=10$, compute $f_1$ or $f_2$
numerically and plot it as a function of M and L (by considering
the positivity condition of blackhole temperature, $-ML^2+2 r_H^3+
r_HL^2 \geq0)$. Then we will have a surface, in the space of
($M,L,f_1$) or ($M,L,f_2$) . Where this surface intersects with
the plane $f_1=0$ or $f_2=0$, is the critical curve of ($M_c ,
L_c$). If there is no intersection then there is no marginally
stable mode for the considered value of $r_H$ with the desired
boundary condition at infinity. Our calculations show that for
\textit{sufficiently large radius of the blackhole} such an
intersection exists.

In figure (1) we have shown focused plots of $f_1$ for two values
of $r_H$. For $r_H=10$ there is no intersection, but, for
$r_H=100$ we have an intersection.

\begin{figure}[H]
  \centering
  \subfloat[]{\includegraphics[scale=.55]{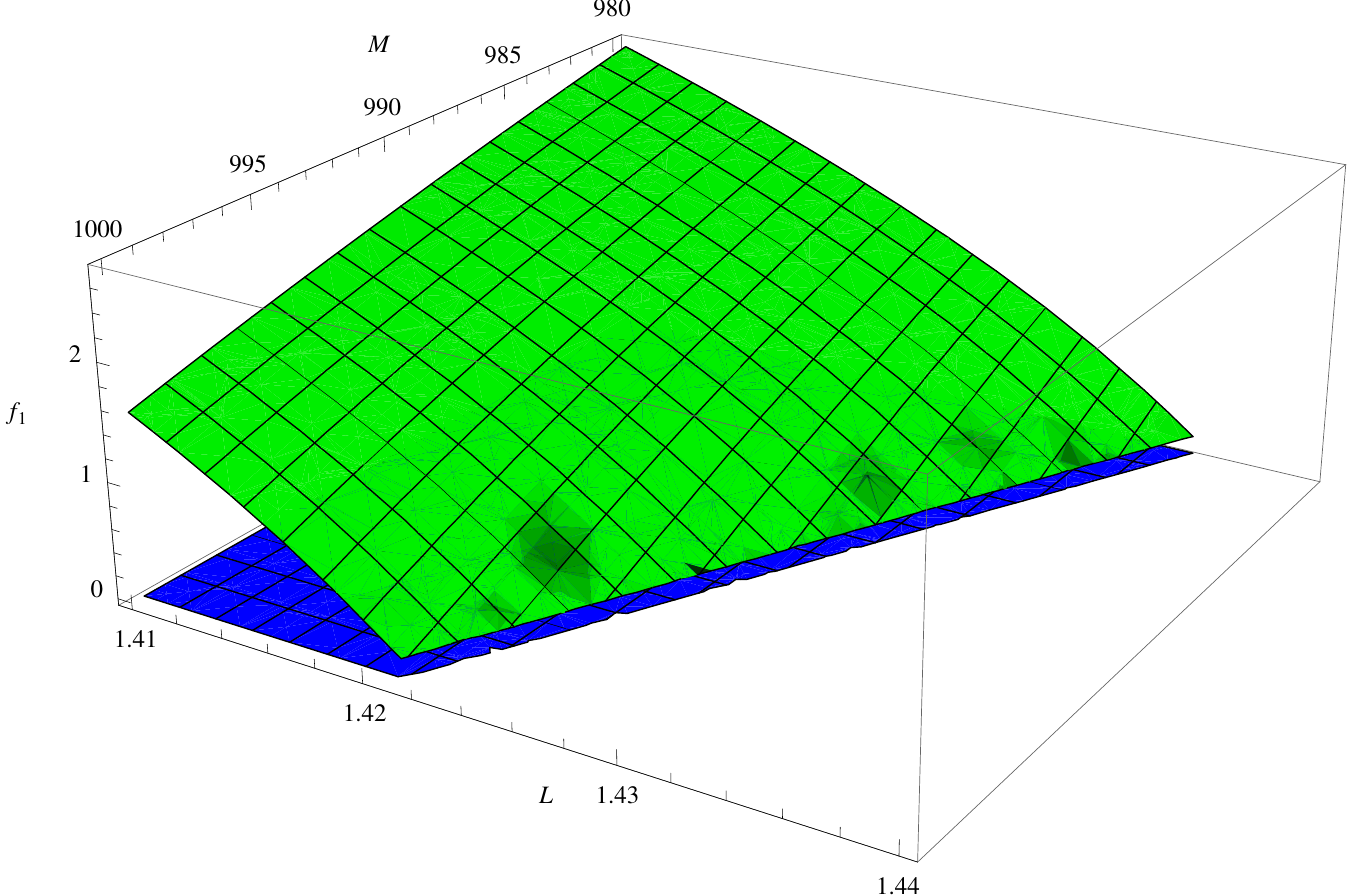}}
  \subfloat[]{\includegraphics[scale=.5]{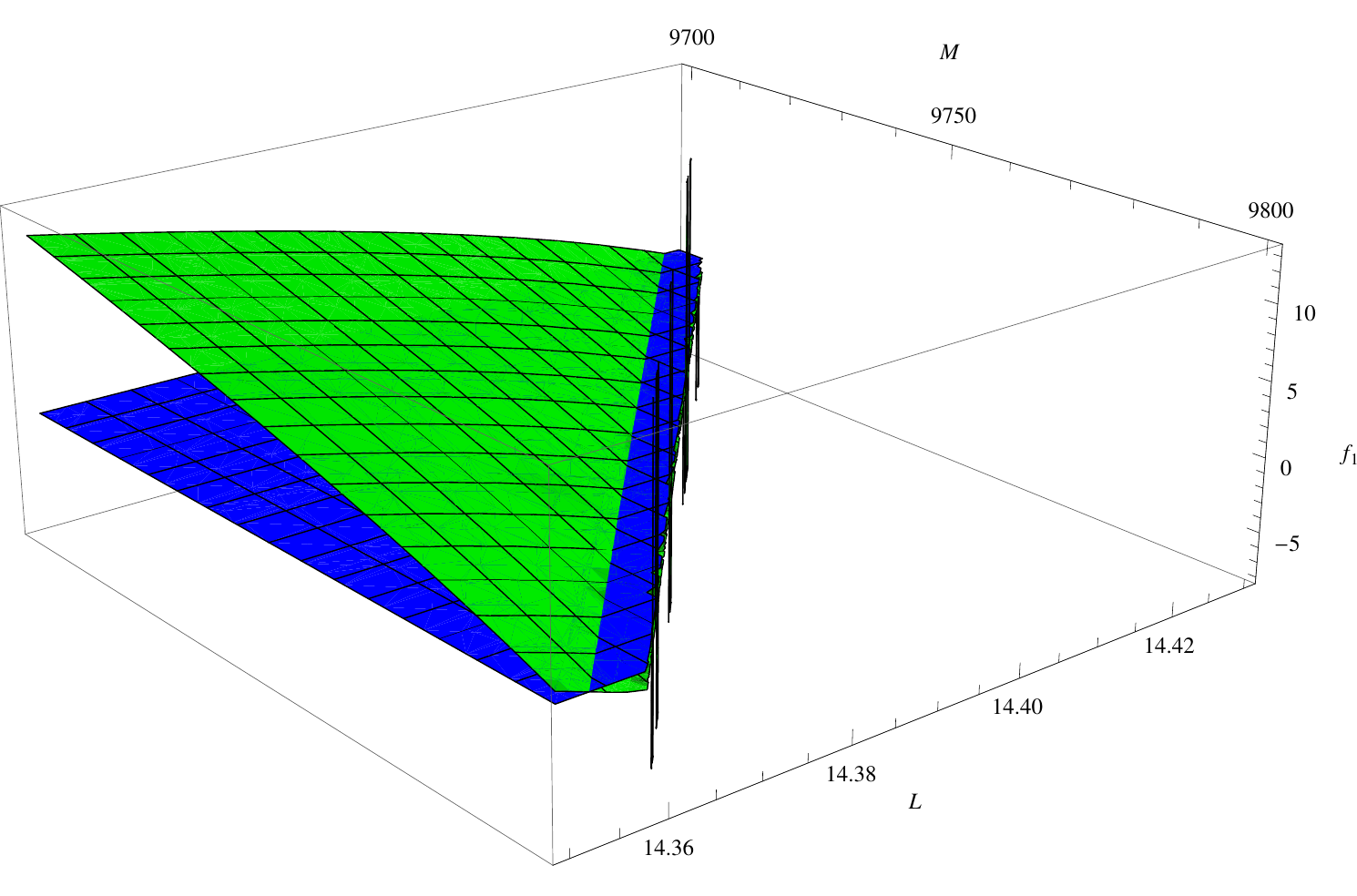}}
  \caption{The green surfaces are the focused plots of $f_1$ as a function of $M$ and
   $L$ and the blue planes are the planes for which $f_1$ vanishes ( plot (a) for $r_H=10$ and plot (b) for $r_H=100$). The plots are cut by $T\geq0$ condition.}
\end{figure}

It can be seen that $f_1$ and $f_2$ can vanish several times
before the temperature becomes zero.

\subsubsection{Quasi-normal modes:}
In this section we introduce our numerical results for
quasi-normal modes. Our aim is to solve the first equation of
(5.4) with the desired boundary conditions and find the complex
frequencies. Our numerical calculations show that at high
temperatures the $\phi=0$ background is stable, the imaginary part
of the quasi-normal frequencies are negative; and at sufficiently
low temperatures this background is unstable, the imaginary part
of the quasi-normal frequencies are positive.

It is obvious that we can not sweep the entire complex frequency
plane by numerical calculations but it seems that at sufficiently
high temperatures there is only one stable mode for every given
temperature (or at least the frequencies in complex frequency
plane are so widely distributed that we could only find one of
them); but, at lower temperatures there are a number of unstable
modes for every value of temperature. The quasi-normal frequencies
that we found are listed in the tables below for both $f_1=0$ and
$f_2=0$.

Note that, specially for $f_1=0$, in a range of temperatures above
the onset of instability, the quasi-normal frequencies are purely
imaginary or their real part is very small; this may be the sign
of the existence of a mass gap right above the critical
temperature $T_c$.

Note also that for $f_1=0$, although there is no marginally stable
mode at $r_H=10$, as mentioned in previous subsection, the
quasi-normal frequencies exhibit a phase transition. This means
that for the boundary conditions under consideration, when the
temperature is changed, the frequencies do not go through the
origin of the complex frequency plane.

\begin{table}[H]
\caption{quasi-normal modes for $f_1=0$} \centering
\begin{tabular}{|c|c|c|}
  \hline
  \multicolumn{3}{|c|}{ $r_H=10$ , $L=1$ } \\
  \hline\hline
  M & $4\pi$ T & $\omega$ \\
  \hline
  2000 & 0.212 & 0.764969 + 1.66858 i \\
  \hline
  2000 & 0.212 & 4.93743 + 3.35654 i \\
  \hline
  1970 & 0.812 & 10.133 + 36.3126 i \\
  \hline
  1970 & 0.812 & 0.669683 + 9.05345 i \\
  \hline
  1950 & 1.212 & 9.9903 + 34.2773 i \\
  \hline
  1900 & 2.211 & 0.0148391 + 35.9313 i \\
  \hline
  1850 & 3.211 & 12.5418 + 52.0683 i \\
  \hline
  1800 & 4.210 & $4.98854\; \times \; 10^{-15}$ - 1.30046 i \\
  \hline
  1700 & 6.210 & $5.26857\;\times\;10^{-14}$ - 2.09467 i \\
  \hline
  1600 & 8.209 & $2.36149\;\times\;10^{-14}$ - 3.01488 i \\
  \hline
  1500 & 10.21 & $1.26155\;\times\;10^{-9}$ - 4.13518 i \\
  \hline
  1200 & 16.21 & 2.16507 - 7.78775 i \\
  \hline
  1200 & 16.21 & 8.66376 - 7.8904 i \\
  \hline
  1000 & 20.20 & 3.7928 - 9.49205 i \\
  \hline
\end{tabular}
\end{table}
\begin{table}[H]
\caption{quasi-normal modes for $f_2=0$}\centering
\begin{tabular}{|c|c|c|}
  \hline
  \multicolumn{3}{|c|}{ $r_H=10$ , $L=1$ } \\
  \hline\hline
  M & $4\pi$ T & $\omega$ \\
  \hline
  2000 & 0.212 & 1.06157 + 1.95297 i \\
  \hline
  1990 & 0.412 & 1.09226 + 3.81241 i \\
  \hline
  1980 & 0.612 & 1.0839 + 5.7522 i \\
  \hline
  1950 & 1.212 & 3.31773 + 19.6673 i \\
  \hline
  1900 & 2.211 & $1.77525\;\times\;10^{-6}$ + 0.320354 i \\
  \hline
  1500 & 10.21 & 0.0000327886 - 0.968159 i \\
  \hline
  1400 & 12.207 & 0.0000337517 - 1.45497 i \\
  \hline
  1000 & 20.20 & 1.55273 - 5.85023 i \\
  \hline
  800 & 24.20 & 4.1098 - 5.3992 i \\
  \hline
\end{tabular}
\end{table}

\subsection{The case $\phi=\sqrt{\frac{2}{\lambda L^2}}$ :}

We have also examined marginally stable modes for The
Kasuya-Kamata solution (3.7). In this case the equations for $f$
when $\omega$ is set to zero, are,\bea
\begin{split} &full \;equation\;
:\;&&\big(r^2e^X\;f'\big)'-\frac{4}{L^2}\;r^2\;f=0,\cr &near
\;horizon\; equation\;
:&&f''+\frac{f'}{u}-\frac{\tilde{\eta}}{u}f=0,\cr
&boundary\;at\;\infty\;equation\;:\;&&f''+\frac{4}{r}f'-\frac{4\alpha^2}{r^2}f=0.\end{split}\eea
The solutions to the near horizon equation are:\be
I_0(2\sqrt{\tilde{\eta}\;u})\; ;\;
K_0(2\sqrt{\tilde{\eta}\;u}).\ee The second solution is divergent
at the horizon so we choose the first one, which has the expansion
: \be 1+\tilde{\eta}\;u\;+\;...\;.\ee The solution to the boundary
equation at infinity is:\bea\begin{split} f\;=
\;f_1\;r^{\Delta_+}\;+\;f_2\;r^{\Delta_-},\cr
\Delta_\pm=\frac{-3\pm\sqrt{9+16\alpha^2}}{2}.\end{split}\eea The
first term is again not acceptable so the second term has to be
chosen.

Finally we should find the appropriate set of parameters (
temperature ) for which the full equation admits a solution with
the desired boundary conditions; and follow the procedure of the
previous section, fixing the horizon condition and searching for
parameters for which $f$ vanishes at the boundary at infinity.

In figure (2) we show the plot of $f$ at large r. We find that it
does not vanish anywhere; therefore the equation does not admit a
solution with the desired boundary condition at any temperature.
Thus when $(\vec{\phi})^2=const\neq0$, the dyon solution, has no
marginally stable mode, indicating stability, in accordance to its
topological nature.

\begin{figure}[H]
  \centering
  \includegraphics[scale=.6]{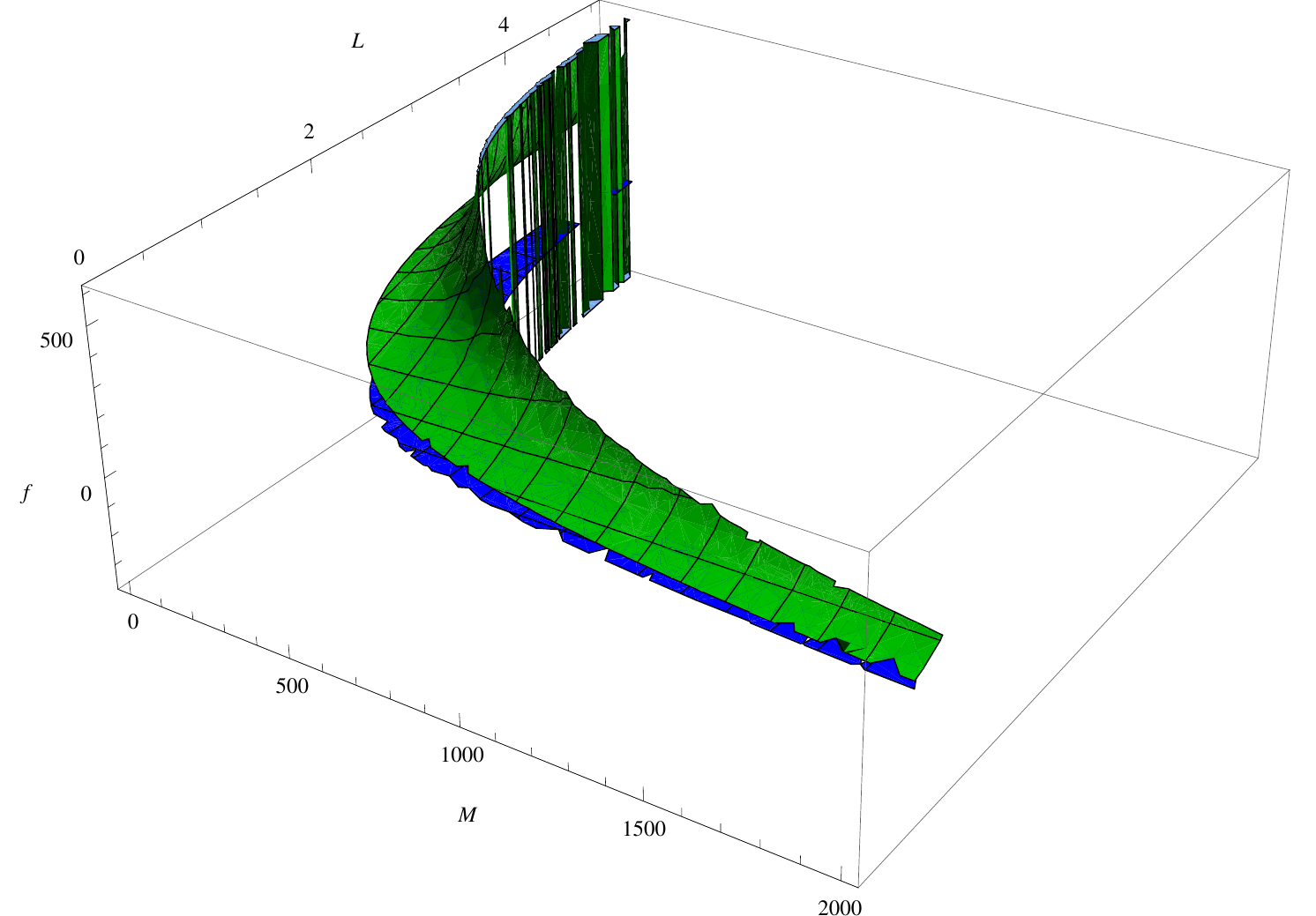}\\
  \caption{The green surface is the plot of $f$ at very large r as a function of $M$ and $L$; and the blue plane is where f vanishes. The plot is cut by $T\geq0$ and $q^2\geq0$ conditions.}
\end{figure}

\section{Conclusion}

In this work we studied stability of two simple dyon solutions of
SU(2) gauge theory in an asymptotically AdS blackhole background
and found that the  solution without the scalar field is unstable.
In the context of holography this is a phase transition for
boundary theory. The solution with the non-vanishing scalar,
describing a topologically nontrivial dyon, turned out to be
stable as expected.  The SU(2) color symmetry  and the SU(2) space
rotational symmetry of the model is broken to a diagonal SU(2)
symmetry in both phases. However the phase transition has a
topological character.

It is tempting to relate this holographic set up to a
gravitational dual of a strong interaction physics model in 2+1
dimensions; the scalar iso-vector field of the model as pions
which are believed to be Goldstone bosons of the spontaneously
broken flavor symmetry.  The color-spin locking in the dyon
configuration is reminiscent of the color-flavor locking in QCD
which is related to color superconductivity in neutron stars. And
of course the intriguing possibility of associating the
topological magnetic configuration with the phenomenon of
confinement in QCD.

It would require further detailed study of the model to make any
firm statements on these issues and we hope to get back to them.

\section{Acknowledgement}
We would like to thank Mohsen Alishahiha, and D.A wants to thank
Ali Davody, Ali Naseh and Hamid Reza Afshar, for useful
discussions.

\end{document}